\documentclass{elsart1p}

\usepackage{graphicx}

\usepackage{amssymb, amsmath}
\begin{document}
\begin{frontmatter}

\title{The Scalars $f_0(980)$ and $a_0(980)$ as Hadronic Molecules}

\author{Tanja Branz\corauthref{cor1}},
\corauth[cor1]{Corresponding author.}
\ead{tanja.branz@uni-tuebingen.de}
\author{Thomas Gutsche, Valery Lyubovitskij\thanksref{label2}}
\thanks[label2]{On leave of absence from the
Department of Physics, Tomsk State University,
634050 Tomsk, Russia}
\address{Institut f\"ur Theoretische Physik, Universit\"at T\"ubingen\\
Auf der Morgenstelle 14, D-72076 T\"ubingen, Germany}

\begin{abstract}
We discuss the hadronic molecule issue in the light and heavy meson sector. Thereby we use the radiative decays of the scalars $f_0(980)$ and $a_0(980)$ to study its possible molecular $K\bar K$ structure. Further on we extend our formalism to mesons with open charm and strangeness, $D^\ast_{s0}(2317)$ and $D_{s1}(2460)$, whose hadronic molecule interpretation is analyzed in weak decays with $f_0(980)$ in the final state. 
\end{abstract}
\end{frontmatter}

\section{Introduction}
There is an ongoing discussion on the structure issue of mesons such as for instance the possible existence of hadronic molecules in the meson sector. In particular mesons lying slightly below a threshold are good candidates for meson-meson bound states. In the following we use a framework which allows for a consistent and fully gauge invariant analysis of hadronic bound states. In addition, the method considers finite size effects which arise due to the extended structure of hadronic molecules.

In the first part we concentrate on the radiative decays of the scalars $f_0(980)$ and $a_0(980)$ within a $K\bar K$ bound state assignment ~\cite{Branz:2008cb}. In the second part we extend our calculations by possible candidates for a molecular structure with open charm and strangeness which are the scalar $D_{s0}^\ast(2317)$ -a possible $DK$ bound state- and the axial $D_{s1}(2460)$ lying close to $D^\ast K$ threshold. Here we study the weak decays with a second hadronic molecule in the final state, the $f_0(980)$~\cite{Branz:2008ha}.

\section{Theoretical framework}
In the present approach mesons lying close to a threshold are assumed to be pure hadronic molecules. In particular we assign $S=K\bar K$ ($S=f_0,a_0$), $D_{s0}^\ast=DK$ and $D_{s1}=D^\ast K$, where $K,D$ and $D^\ast$ represent the isospin doublets (see \cite{Faessler:2007us}).

The basic element of our model is provided by the nonlocal Lagrangian describing the coupling of the hadronic molecule $H$ to the respective constituent mesons $M_1$ and $M_2$\\\vspace{-0.3cm}
\begin{equation}
{{\cal L}_{HM_1M_2}= g_{H} H(x) \int dy 
\,\Phi_H(y^2) \, M_1(x+w_{21}y) \, M_2(x+w_{12}y)} + \text{H.c.}\label{eq:L}
\end{equation}
with $w_{ij}=\frac{m_i}{m_i+m_j}$ and the coupling constant $g_H$.

Finite size effects are considered by the vertex function $\Phi(y^2)=\int\frac{d^4k}{(2\pi)^4}e^{-iky}\widetilde \Phi(-k^2)$. In momentum space we use a Gaussian form $\widetilde\Phi(k_E^2)=\exp(-k_E^2/\Lambda^2)$ with the size parameter $\Lambda$ = 0.7 - 1.3 GeV. Provided that the decay amplitude is free of UV divergences we also compute the local limit, where $\Lambda\to\infty$ simulates point-like interaction.

A self-consistent description of bound states is given by the compositeness condition \cite{Weinberg:1962hj}, where we realize a composite object by setting its renormalization constant $Z_{H}$ to zero. $Z_{f_0}=1-g_H^2\tilde \Pi^{\prime}(m^2_{H})=0$ is characterized by the derivative of the mass operator $\Pi(m^2_H)$ allowing for a consistent determination of the coupling $g_H$.

\section{Radiative decays}
\begin{figure}[thbp]
\begin{center}
\vspace*{-0.05cm}
\includegraphics*[trim= 0cm 0cm 0cm 0cm,clip, scale=0.38]{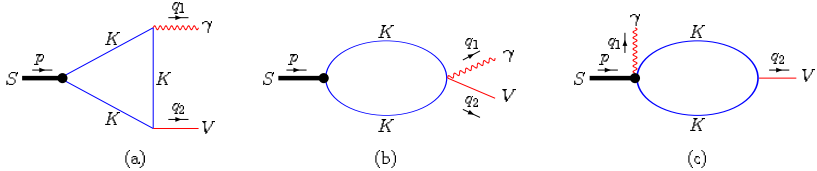}
\end{center}
\caption{Radiative decays of $S\to V\gamma$ with ($S=f_0(980),a_0(980)$, $V=\rho,\omega$).}
\label{fig:1}
\end{figure}

Electromagnetic interaction is included by minimal substitution. In case of nonlocal Lagrangians (\ref{eq:L}) we write $M_{1,2}^{\pm}(y)\rightarrow \exp\Big[\mp ie\int\limits_x^ydz_\mu A^\mu(z)\Big]M_{1,2}^\pm(y)$~\cite{Terning:1991yt}
 which results in additional diagrams contributing to the radiative decays (see e.g. Fig. \ref{fig:1}~c)). In the hadronic molecule picture all decays proceed via intermediate constituents, here by kaon loops.

The coupling of the kaons to the vectors $V=\rho,\omega$ is given by the standard Lagrangian\\
${\cal L}_{VK\bar K}=g_{\rho K\bar K}\vec \rho^{\,\mu}\bar K \vec\tau \,i\partial_\mu K+\!\!\!\!\sum\limits_{V=\phi,\omega}\!\!\!g_{V K\bar K}V^\mu\bar K  \,i\partial_\mu K+h.c.$ 
with $g_{\rho K\bar K}=g_{\omega K\bar K}=\frac{g_{\phi K\bar K}}{\sqrt2}=3$ fixed by $SU(3)$ symmetry relations. Our results for the two photon decays and the decays with a final vector meson are summarized in Tabs.~\ref{tab:1} and~\ref{tab:2}.
\def\arraystretch{1.0} 
\begin{table}[htbp]
\begin{minipage}[t]{8.9cm}
\caption{Decay widths $\Gamma(S\to\gamma\gamma)$, ($S=f_0,a_0$) in keV.}\label{tab:1}
\begin{tabular}{|llc|cll|}
\hline
\multicolumn{2}{|c}{$\Gamma(f_0\to\gamma\gamma)$}&&&\multicolumn{2}{c|}{$\Gamma(a_0\to\gamma\gamma)$}\\\hline
Data \cite{Amsler:2008zz}&$0.29^{+0.07}_{-0.09}\;$&&&Data \cite{Amsler:1997up}&$0.3\pm0.1\;$\\
Our ($\Lambda$=1 GeV)$\quad$&0.25&&&Theo. ($\Lambda$=1 GeV)$\quad$&0.19\\
Our (local)&0.29&&&Theo. (local)&0.23\\\hline
\end{tabular}
\end{minipage}\hfill
\begin{minipage}[t]{6cm}
\caption{Decay widths $\Gamma(S\to V\gamma)$ in keV.}\label{tab:2}
\begin{tabular}{|lll|}
\hline
Decay mode$\quad$ &nonlocal$\quad$&local\\\hline
$\Gamma(f_0\to\rho\gamma)$&7.58 &8.09\\
$\Gamma(f_0\to\omega\gamma)$&7.12 &7.57 \\
$\Gamma(a_0\to\rho\gamma)$&6.59 &7.18\\
$\Gamma(a_0\to\omega\gamma)$&6.22& 6.76\\\hline
\end{tabular}
\end{minipage}
\end{table}
For the production of $f_0$ and $a_0$ in $\phi$ decays we obtain $\Gamma(\phi\to f_0\gamma)=0.64$ keV and $\Gamma(\phi\to a_0\gamma)=0.42$ keV.

\section{Weak decays}

\begin{figure}[thbp]
\begin{center}
\includegraphics*[trim= 0cm 1cm 0cm 0cm,clip, scale=0.4]{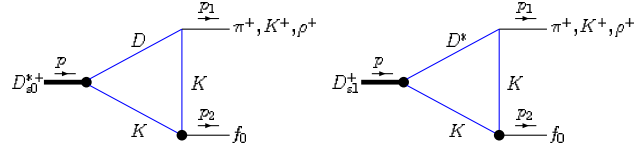}
\end{center}
\caption{$D_{s0}^\ast$ and $D_{s1}$ weak decay processes.}
\label{fig:2}
\end{figure}

Besides the light meson sector the present framework can also be applied to heavy mesonic bound states such as the $D_{s0}^\ast$ and $D_{s1}$. The weak decays $D_{s0}^\ast\to f_0X$ and $D_{s1}\to f_0X$ proceed via the intermediate constituents of the respective final and initial hadronic molecules (see Fig.~\ref{fig:2}). The couplings are either taken from experiment or, in the case of hadronic molecules, fixed by the compositeness condition.
The results for the respective $D_{s0}^\ast$ decay modes differ by an order of magnitude as indicated in Tab.~\ref{tab:3}.
Together with $\Gamma(D_{s1}\to f_0\pi)$ = (2.85 - 4.35)$\cdot10^{-11}$ GeV ($\Lambda_D$= 1 - 2 GeV)
 the weak decays span a region of four orders of magnitude.
\begin{table}[htbp]
\caption{Decay widths of the weak decays $D_{s0}^{\ast}\to f_0X$ ($X=K,\pi,\rho$) in GeV.}\label{tab:3}
\begin{tabular}{|lll|}
\hline
&nonlocal$\qquad$&$\qquad$local\\\hline
$\Gamma(D_{s0}^{\ast}\to f_0K)$$\qquad$  &(1.42 - 1.53)$\cdot10^{-15}$ &$\qquad$$2.75\cdot10^{-15}$\\
$\Gamma(D_{s0}^{\ast}\to f_0\pi)$&(1.14 - 1.26)$\cdot10^{-14}$&$\qquad$$2.35\cdot 10^{-14}$ \\
$\Gamma(D_{s0}^{\ast}\to f_0\rho)$&(1.08 - 1.11)$\cdot 10^{-13}$&$\qquad$$1.60\cdot 10^{-13}$\\\hline
\end{tabular}
\end{table}
\section{Conclusions}

The present QFT approach for hadronic molecules based on the Weinberg condition obeys Lorentz covariance and gauge invariance. It provides a consistent determination of production and decay properties. The model includes finite size effects controlled by the size parameters $\Lambda$. The electromagnetic decay and production properties are in good agreement with data. Further on we extended our formalism to weak transitions between hadronic molecules which could possibly test the molecular structure as well. \\

{\bf{Acknowledgments}}\\

This work was supported by the DFG under Contract No. FA67/31-1,
No. FA67/31-2, and No. GRK683. This research is also part of the
European Community-Research Infrastructure Integrating Activity
``Study of Strongly Interacting Matter'' (HadronPhysics2,
Grant Agreement No. 227431) and of the President grant of Russia
``Scientific Schools''  No. 871.2008.2.
T.B. would like to thank the Organizing Committee of the
PANIC 08 for the financial support.


\begin{thebibliography}{00}
%
\bibitem{Branz:2008cb}
  T.~Branz, T.~Gutsche and V.~E.~Lyubovitskij,
  Phys.\ Rev.\  D {\bf 79}, 014035 (2009)


\bibitem{Branz:2008ha}
  T.~Branz, T.~Gutsche and V.~E.~Lyubovitskij,
  Phys.\ Rev.\  D {\bf 78}, 114004 (2008); 
  T.~Branz, T.~Gutsche and V.~E.~Lyubovitskij,
  Eur.\ Phys.\ J.\  A {\bf 37}, 303 (2008).

\bibitem{Faessler:2007us}
  A.~Faessler, T.~Gutsche, V.~E.~Lyubovitskij and Y.~L.~Ma,
  Phys.\ Rev.\  D {\bf 76}, 114008 (2007); 
  
  A.~Faessler, T.~Gutsche, V.~E.~Lyubovitskij and Y.~L.~Ma,
  Phys.\ Rev.\  D {\bf 76}, 014005 (2007).




\bibitem{Weinberg:1962hj}
  S.~Weinberg,
  Phys.\ Rev.\  {\bf 130}, 776 (1963); 
  A.~Salam,
  Nuovo Cim.\  {\bf 25}, 224 (1962).


\bibitem{Terning:1991yt}
  J.~Terning,
  Phys.\ Rev.\  D {\bf 44}, 887 (1991).

\bibitem{Amsler:2008zz}
  C.~Amsler {\it et al.}  [Particle Data Group],
  Phys.\ Lett.\  B {\bf 667}, 1 (2008).

\bibitem{Amsler:1997up}
  C.~Amsler,
  Rev.\ Mod.\ Phys.\  {\bf 70}, 1293 (1998)

%
\end{thebibliography}
\end{document}